\documentclass[prl,preprint,superscriptaddress,showpacs,showkeys,amsmath,amssymb,aps]{revtex4-1}

\setcitestyle{super}
\usepackage{graphicx}
\usepackage[utf8]{inputenc}
\usepackage{amsmath}
\usepackage[version=3]{mhchem}
\usepackage[usenames]{color}
\usepackage{color}
\usepackage{physics}
\usepackage[colorlinks=true, linkcolor=blue, citecolor=blue, urlcolor=blue,pdftex]{hyperref}
\definecolor{newcolor}{rgb}{0.9,0,0.1}

\usepackage{orcidlink}
\usepackage{multirow}
\usepackage{siunitx}
\usepackage{float}
\usepackage{svg}

\newcommand{\cminv}{cm\textsuperscript{-1}}
\newcommand{\wse}{WSe\textsubscript{2}}
\newcommand{\sio}{SiO\textsubscript{2}}

\usepackage{xr}
\makeatletter
\newcommand*{\addFileDependency}[1]{
\typeout{(#1)}
\@addtofilelist{#1}
\IfFileExists{#1}{}{\typeout{No file #1.}}

\makeatother
}

\begin{document}

\title{Layer-Dependent Interfacial Coupling and Exciton Pinning in \wse{}/Graphene Heterostructures}

\author{Lan Huang\,\orcidlink{} \textsuperscript{\textdagger}}
\affiliation{Transport at Nanoscale Interfaces Laboratory, Empa -- Swiss Federal Laboratories for Materials Science and Technology, D\"ubendorf 8600, Switzerland}

\author{Laric Bobzien\,\orcidlink{0009-0002-9792-3567} \textsuperscript{\textdagger}}
\affiliation{nanotech@surfaces Laboratory, Empa -- Swiss Federal Laboratories for Materials Science and Technology, D\"ubendorf 8600, Switzerland}

\author{Ángel Labordet Álvarez\,\orcidlink{}}
\affiliation{Transport at Nanoscale Interfaces Laboratory, Empa -- Swiss Federal Laboratories for Materials Science and Technology, D\"ubendorf 8600, Switzerland}
\affiliation{Department of Physics, University of Basel, Basel, 4056 Switzerland}
\affiliation{Swiss Nanoscience Institute, University of Basel, Basel, 4056 Switzerland}

\author{Daniel E. Cintron Figueroa}
\affiliation{Department of Chemistry, The Pennsylvania State University, University Park, 16802, PA, USA}

\author{Li-Syuan Lu}
\affiliation{Department of Materials Science and Engineering, The Pennsylvania State University, University Park, 16802, PA, USA}

\author{Chengye Dong\,\orcidlink{0000-0001-8598-5758}}
\affiliation{Two-Dimensional Crystal Consortium and Materials Research Institute, The Pennsylvania State University, University Park, 16802, PA, USA}

\author{Michel Calame\,\orcidlink{}}
\affiliation{Transport at Nanoscale Interfaces Laboratory, Empa -- Swiss Federal Laboratories for Materials Science and Technology, D\"ubendorf 8600, Switzerland}
\affiliation{Department of Physics, University of Basel, Basel, 4056 Switzerland}
\affiliation{Swiss Nanoscience Institute, University of Basel, Basel, 4056 Switzerland}

\author{Joshua A. Robinson\,\orcidlink{0000-0002-1513-7187}}
\affiliation{Department of Chemistry, The Pennsylvania State University, University Park, 16802, PA, USA}
\affiliation{Two-Dimensional Crystal Consortium and Materials Research Institute, The Pennsylvania State University, University Park, 16802, PA, USA}
\affiliation{Department of Physics and Department of Materials Science and Engineering, The Pennsylvania State University, University Park, 16802, PA, USA}

\author{Bruno Schuler\,\orcidlink{0000-0002-9641-0340}}

\affiliation{nanotech@surfaces Laboratory, Empa -- Swiss Federal Laboratories for Materials Science and Technology, D\"ubendorf 8600, Switzerland}

\author{Mirjana Dimitrievska\,\orcidlink{0000-0002-9439-1019}}
\email[]{Mirjana.Dimitrievska@empa.ch}
\affiliation{Transport at Nanoscale Interfaces Laboratory, Empa -- Swiss Federal Laboratories for Materials Science and Technology, D\"ubendorf 8600, Switzerland}


\begin{abstract}

Understanding interfacial interactions in two-dimensional heterostructures is crucial for their implementation in future optoelectronic and quantum technologies. Here, we investigate the interactions between \wse{} and graphene by comparing metal-organic chemical vapor deposition (MOCVD)-grown \wse{} films containing 1--5 layers on graphene/SiC with exfoliated \wse{} on \sio{}/Si using Raman and photoluminescence spectroscopy combined with atomic force microscopy. We find that growth on graphene induces persistent compressive strain on the order of 0.2\% and reduces the interlayer \wse{} distance by $0.11 \pm 0.05$~\AA. The degree of interfacial disorder is strongest in the \wse{} layer directly contacting graphene, with an effective Urbach energy of $\sim20$~meV, which decreases to $\sim16$~meV upon addition of a second layer and remains similar in subsequent \wse{} layers. Increasing \wse{} thickness leads to progressive electron transfer from graphene to \wse{}, resulting in $p$-type doping of graphene and $n$-type doping of \wse{}, with the graphene hole density increasing from approximately $0.4\times10^{13}$~cm$^{-2}$ for 1L to $0.8\times10^{13}$~cm$^{-2}$ for 5L. The A- and B-exciton energies of \wse{} remain nearly pinned on graphene, in contrast to their pronounced thickness-dependent shifts on \sio{}/Si. We show that this pinning is primarily due to electronic screening and compressive strain, with smaller contributions from charge transfer and modified interlayer coupling. These findings establish graphene as an active interface for controlling excitonic properties in scalable van der Waals heterostructures.

\noindent\textbf{Highlight:} Graphene acts not merely as a support but as an active interfacial partner, pinning excitonic states in \wse{} and enabling control over optical properties beyond what is possible on conventional dielectric substrates.

\begin{center}
\includegraphics[width=0.75\linewidth]{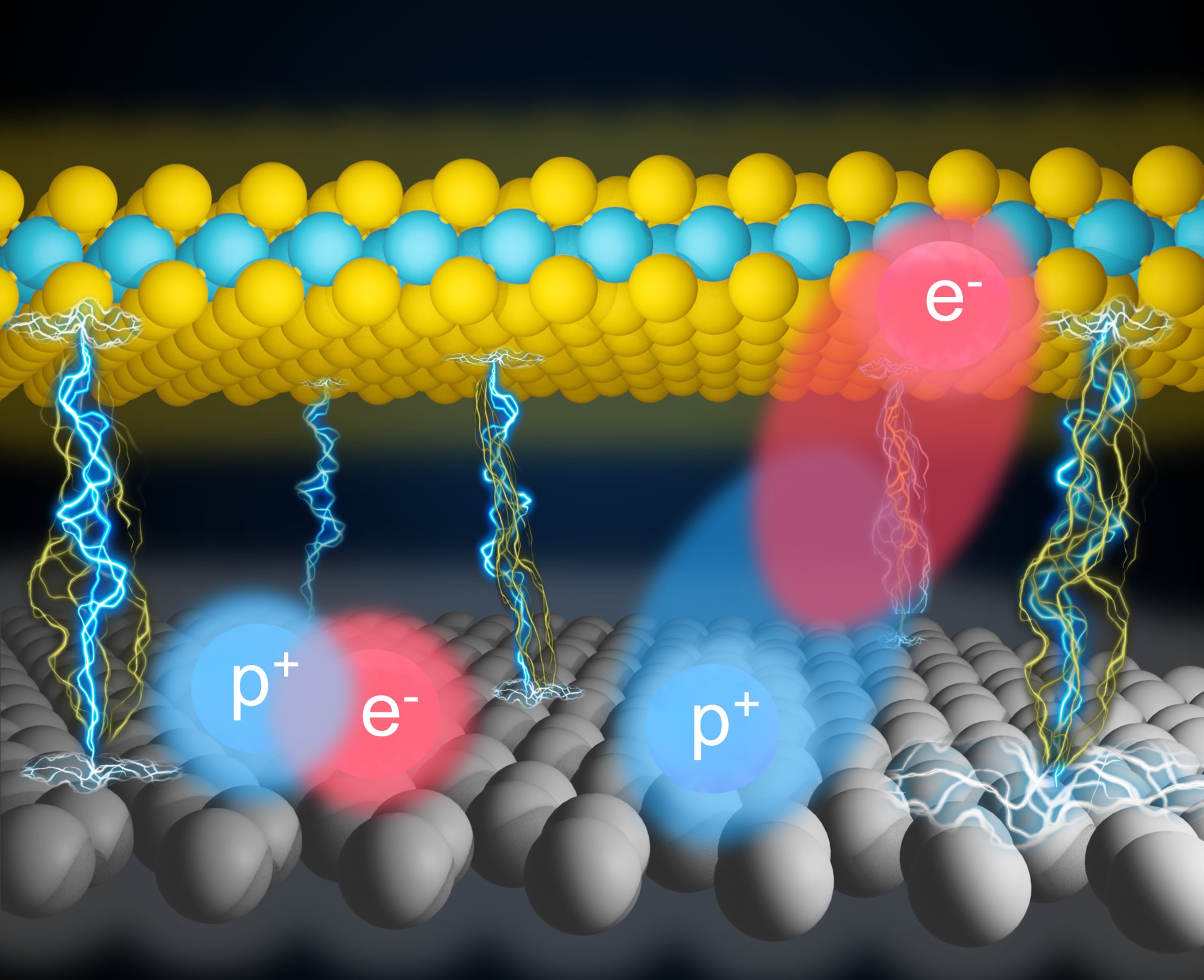}
\end{center}
\end{abstract}


\date{\today}
\maketitle


\section{Introduction}\label{sec1}

Two-dimensional (2D) transition metal dichalcogenides (TMDs), such as tungsten diselenide (\wse{}), have emerged as promising materials for a wide range of optoelectronic and quantum applications.\cite{eftekhari2017tungsten,mak2016photonics,dastidar2022quantum} Their strong light--matter interaction results in large exciton binding energies, while spin--orbit coupling gives rise to valley splitting and the layer-dependent band gap becomes direct in the monolayer limit.\cite{RevModPhys.90.021001} This makes monolayer \wse{} an ideal candidate for light-emitting devices, photodetectors, and valleytronic applications.\cite{ahmed2023bright,shi2025recent,schaibley2016valleytronics} As the number of layers increases, \wse{} retains a high absorption coefficient and develops distinctive interlayer phonon modes, which are relevant for applications in nanoelectronics, sensing, and strain-engineering platforms.\cite{zhang_raman_2013,o2016mapping}

The inherently weak out-of-plane dielectric screening in 2D materials renders their electronic and optical properties exquisitely sensitive to local environmental perturbations and adjacent layers in van der Waals heterostructures.\cite{raja_coulomb_2017} Electrostatic gating offers dynamic control of their properties but requires device fabrication and primarily modifies the carrier density and Fermi level, although band-gap renormalization may also occur at high doping levels.\cite{nguyen2019visualizing} By contrast, tailoring the interface provides a direct route to modify the electronic and optical properties through strain-induced changes in the band gap and phonon spectrum,\cite{aslan2018strain,zhang_raman_2013} interfacial charge transfer,\cite{kim2017observation,jin2018imaging} and environmental screening of the quasiparticle gap and exciton binding energy.\cite{raja_coulomb_2017,borghardt2017engineering} Van der Waals coupling can further modify interlayer phonon behaviour and exciton dynamics.\cite{jin2017interlayer} These effects are particularly relevant when 2D semiconductors are integrated with conductive materials such as graphene. 

Graphene--\wse{} heterostructures have become model systems for exploring interfacial coupling mechanisms in 2D materials.\cite{lorchat2020filtering,froehlicher2018charge} Raman and photoluminescence spectroscopy have been used to examine how graphene modifies the vibrational and optical properties of exfoliated \wse{}. Electrical measurements revealed significant band-edge shifts in graphene--\wse{} heterostructures,\cite{kim2015band} while PL and Raman studies identified emission quenching, ultrafast charge transfer, and strain effects.\cite{rieland_ultrafast_2024,xu_tunable_2022} The proximity of graphene was also shown to suppress PL and modify spin--orbit interactions depending on the interlayer distance.\cite{yang_effect_2018} Scanning tunneling spectroscopy further mapped the thickness-dependent evolution of the band edges in undoped and V-doped \wse{} on epitaxial graphene, showing how band-gap narrowing and Fermi-level pinning develop through interlayer coupling and doping.\cite{stolz_layer-dependent_2022} However, previous far-field studies are based mainly on exfoliated \wse{} placed on graphene, or graphene placed on \wse{}, and focus primarily on how the number of graphene layers affects monolayer or bilayer \wse{}. Systematic studies varying the number of as-grown \wse{} layers to determine how the \wse{}--graphene coupling evolves are still lacking. Moreover, a direct comparison between as-grown \wse{} on graphene and exfoliated \wse{} on \sio{}/Si, the standard substrate for fundamental studies, has not yet been reported. This comparison is particularly relevant because direct growth on graphene/SiC is important for scalable device applications.

In this study, we investigate the coupling between \wse{} and graphene using metal-organic chemical vapor deposition (MOCVD)-grown \wse{} with thicknesses from 1L to 5L on quasi-freestanding epitaxial graphene/SiC, directly compared with exfoliated \wse{} on \sio{}/Si. We evaluate strain, interlayer coupling, and charge transfer by tracking the evolution of the \wse{} intra- and interlayer phonon modes and the graphene G and 2D modes using Raman spectroscopy. We then examine the layer-dependent emission, exciton-energy shifts, and energetic disorder using photoluminescence (PL) spectroscopy. By combining these measurements, we quantify the thickness-dependent doping, compressive strain, and changes in interlayer distance, and assess their contributions to the excitonic response of \wse{}.

\section{Results and Discussion}\label{sec2}

\subsection{Morphological Characterization of MOCVD-Grown \wse{} on Graphene/SiC substrate}

\begin{figure}[t]
\centering
\includegraphics[width=0.7\textwidth]{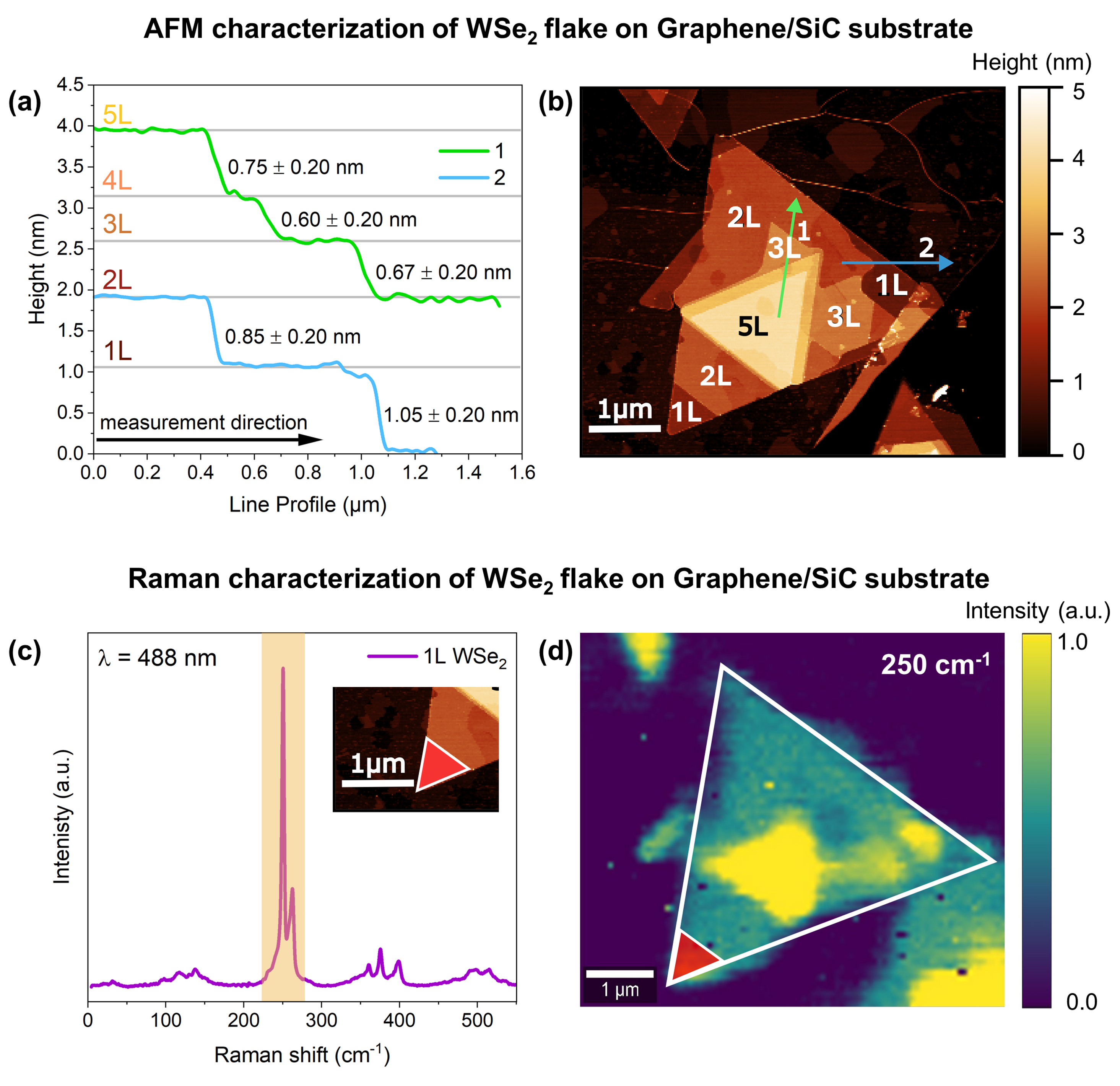}
\caption{\label{fig1}\textbf{Morphological Characterization of MOCVD-Grown \wse{} on Graphene/SiC substrate.} (a) AFM height profiles extracted along the lines indicated in (b), used to assign \wse{} thickness from 1L to 5L. (b) AFM image of the \wse{} flake measured in tapping-mode showing the height profile and confirming the assignment of the different numbers of layers. (c) Representative Raman spectrum of the \wse{} flake measured on the monolayer area indicated by the red triangle in the inset AFM image and in (d), and using an excitation wavelength of 488\,nm. The area around 250\,cm\textsuperscript{-1} is highlighted in the spectrum and its intensity over the flake is shown on the Raman image in (d).}
\end{figure}

We first evaluate the morphological and structural integrity of the MOCVD-grown \wse{}/graphene heterostructures using atomic force microscopy (AFM), supplemented by Raman spectroscopy. Figure~\ref{fig1} presents an overview of a single \wse{} island on bilayer graphene on a SiC substrate. 
Figure~\ref{fig1}a and~\ref{fig1}b show the AFM topography of the as-grown flake, which displays a triangular morphology characteristic of MOCVD-grown \wse{}. The contrast in the height map reveals discrete terraces, which were assigned to monolayer (1L), bilayer (2L), trilayer (3L), four-layer (4L), and five-layer (5L) regions, as labeled. These assignments are supported by the height profiles shown in Figure~\ref{fig1}a, extracted along the marked directions (lines 1 and 2 in the AFM image). The measured step heights between terraces range from 0.60 to 0.85\,nm, consistent with the expected interlayer spacing in few-layer \wse{}. These values fall within the typical range reported for TMDs measured on various substrates and may include contributions from instrumental uncertainty or local variations in layer–substrate interaction.\cite{zhang2019afmlayer,tonndorf2013afmlayer,blaga_unveiling_2024} The lateral uniformity of each layer across several microns enables reliable Raman and PL measurements.

To further confirm the crystal quality of the flake, we performed Raman measurements using a 488\,nm excitation wavelength. A representative Raman spectrum acquired from the monolayer region is shown in Figure~\ref{fig1}c. The dominant feature near 250\,\cminv{} corresponds to the overlapping E\textsubscript{2g} and A\textsubscript{1g} phonon modes of \wse{}, which appear as a single unresolved peak under non-polarized backscattering conditions.\cite{blaga_unveiling_2024,lucaRaman} It should be noted that mode notations differ in multilayer \wse{} systems: the A\textsubscript{1g} and E\textsubscript{2g} modes of the bulk correspond to A\textsubscript{1}' and E' modes for an odd number of layers, and to A\textsubscript{1g} and E\textsubscript{g} modes for an even number of layers. For easier comparison, we adopt the bulk mode notation throughout this study. The absence of low-frequency shear and breathing modes below 20\,\cminv{} confirms the monolayer nature of this region.\cite{blaga_unveiling_2024,zhao_interlayer_2013} Multi-phonon modes are also observed at higher Raman frequencies ($>$310\,\cminv{}), which are characteristic of high-crystal-quality \wse{}.\cite{blaga_unveiling_2024,lucaRaman}
Figure~\ref{fig1}d displays a Raman intensity map of the 250\,\cminv{} peak over the flake area. The triangular flake contour, overlaid in white, matches the morphology observed in AFM. The map shows an overall consistent signal intensity within each terrace, indicative of uniform crystallinity. When comparing different layer thicknesses, the intensity of the 250\,\cminv{} peak follows the expected trend for 488\,nm excitation: it is strongest for thicker layers, and less intense for 1L and 2L, consistent with resonance behavior and exciton–phonon coupling reported in the literature.\cite{blaga_unveiling_2024}

\subsection{Raman Characterization of MOCVD-Grown \wse{} on Graphene/SiC substrate}\label{sec3}

\begin{figure}[t]
\centering
\includegraphics[width=0.65\textwidth]{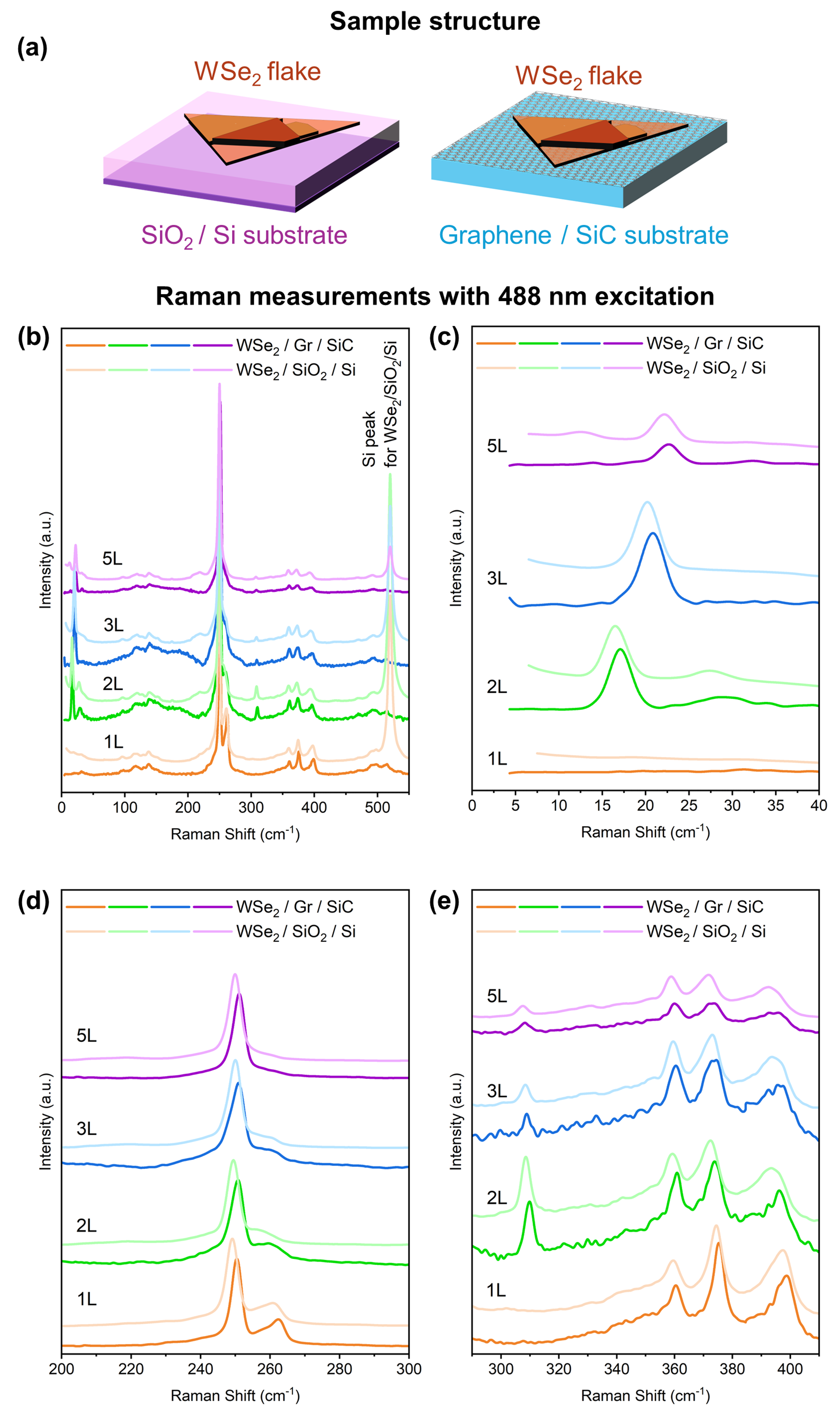}
\caption{\label{fig2}\textbf{Raman characterization of \wse{} flakes on graphene/SiC and SiO\textsubscript{2}/Si substrates using 488\,nm excitation.} 
(a) Sample structures: MOCVD-grown \wse{} on bilayer graphene/SiC and exfoliated \wse{} on \sio{}/Si. (b) Comparison of the Raman spectra for 1L to 5L of MOCVD-grown \wse{} on bilayer graphene/SiC and exfoliated \wse{} on \sio{}/Si. (c–e) Low-, mid-, and high-frequency Raman regions, showing interlayer shear and breathing modes (c), the A\textsubscript{1g}/E\textsubscript{2g} phonon mode near 250,\cminv{} (d), and higher-order phonon modes  above 300\,\,\cminv{} (e). All spectra were normalized to the A\textsubscript{1g}/E\textsubscript{2g} mode for comparison.}
\end{figure}

To investigate the vibrational properties and interfacial effects in \wse{} grown on graphene, we perform Raman measurements using 488\,nm excitation and compare the results to our reference dataset of exfoliated \wse{} on \sio{}/Si, as we reported previously in Blaga \textit{et al.}\cite{blaga_unveiling_2024}. Figure~\ref{fig2} presents the Raman spectra of \wse{} flakes with layer numbers ranging from monolayer (1L) to five layers (5L) for both sample types. As shown in Figure~\ref{fig2}a, the MOCVD-grown \wse{} is supported on bilayer graphene/SiC, while the exfoliated flakes are placed on thermally grown \sio{}/Si. Figure~\ref{fig2}b--e include both full-range Raman spectra and zoomed-in views of specific frequency windows, highlighting the low-frequency interlayer modes, the main A\textsubscript{1g}/E\textsubscript{2g} mode around 250\,\cminv{}, and higher-order modes beyond 300\,\cminv{}. Note that each spectrum represents an average of all spectra acquired within the corresponding layer region identified in the AFM image in Figure~\ref{fig1}b, as detailed in the Experimental Methods section. All spectra have been normalized to the A\textsubscript{1g}/E\textsubscript{2g} mode for comparison.

The full-range spectra in Figure~\ref{fig2}b show that both sample types have the expected phonon modes of \wse{}, with prominent peaks around 250\,\cminv{} and additional features near 260, 309, and 375\,\cminv{}, which correspond to the 2LA(M), B2g, and multi-phonon modes, respectively.\cite{blaga_unveiling_2024,lucaRaman} When comparing the Raman intensities, we observe that for \wse{} on \sio{}, the absolute intensity is higher, as seen from the higher signal to noise ratio compared to the \wse{} on graphene. This is in agreement with the stronger Raman response typically observed on dielectric substrates due to reduced quenching and enhanced light–matter interaction.\cite{li2012RamanInt,lien2015RamanInt,buscema2014RamanInt} In contrast, \wse{} on graphene exhibits systematically lower Raman intensities, which we attribute to graphene’s light absorption and its ability to facilitate faster carrier relaxation and non-radiative processes.\cite{yang_effect_2018,lorchat2020filtering}
Figure~\ref{fig2}c focuses on the low-frequency region, where interlayer shear and breathing modes become visible starting from bilayer \wse{}. The absence of low-frequency modes in the monolayer and their progressive appearance and shift with increasing thickness are consistent with the expected behavior for van der Waals layered materials \cite{zhao_interlayer_2013}. The layer-dependent evolution of these low-frequency modes is further detailed in Figure~\ref{fig4}, where we show the extracted Raman frequencies as a function of layer number.

\begin{figure}[t]
\centering
\includegraphics[width=1.0\textwidth]{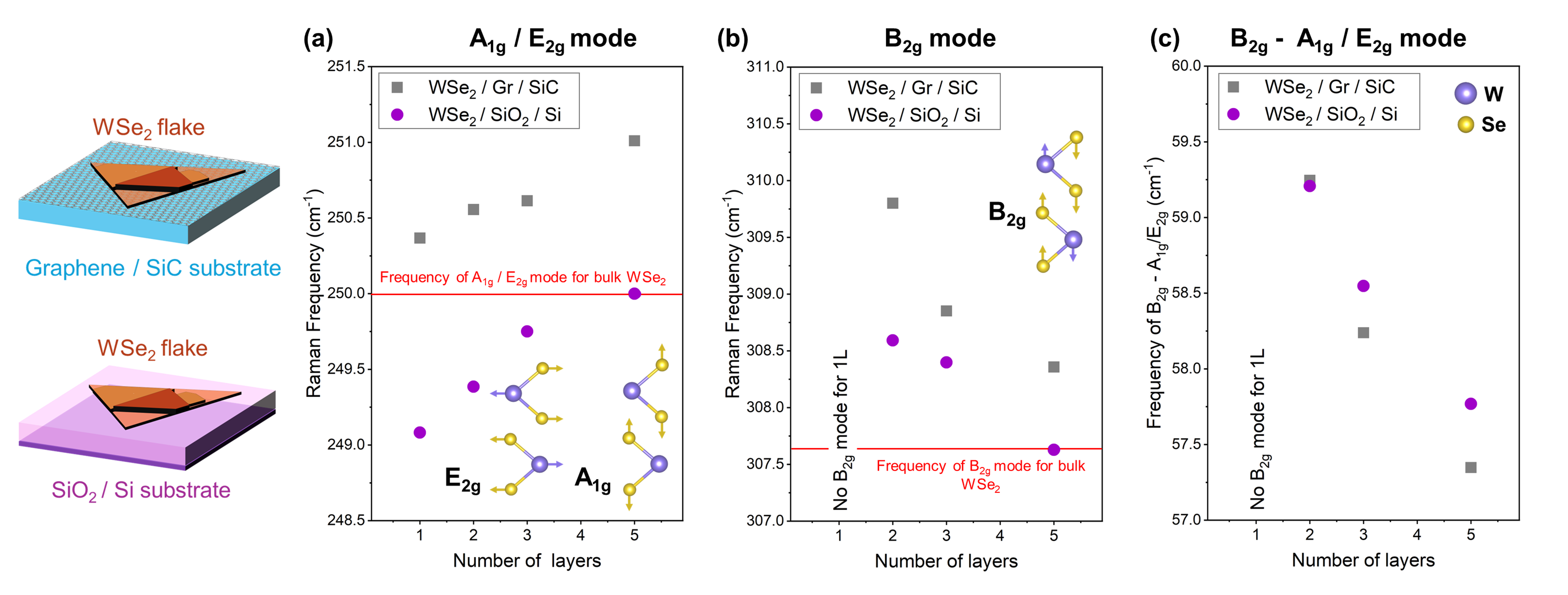}
\caption{\label{fig3} \textbf{Layer-dependent Raman analysis of \wse{} on graphene/SiC and \sio{}/Si substrates.}
(a) A\textsubscript{1g}/E\textsubscript{2g} and (b) B\textsubscript{2g} mode frequencies and (c) their difference (B\textsubscript{2g} - A\textsubscript{1g}/E\textsubscript{2g}) as a function of \wse{} layer thickness on graphene/SiC (gray squares) and on \sio{}/Si (purple circles from \cite{blaga_unveiling_2024}). Red lines indicate the frequencies of A\textsubscript{1g}/E\textsubscript{2g} and  B\textsubscript{2g} modes for bulk \wse{}. }
\end{figure}

\subsubsection{Intralayer Phonon Modes of \wse{}}

Figure~\ref{fig3}a and~\ref{fig3}b present the frequencies of the A\textsubscript{1g}/E\textsubscript{2g} and B\textsubscript{2g} modes for both substrate types. Previous studies have shown that the A\textsubscript{1g}/E\textsubscript{2g} mode shows higher sensitivity to in-plane strain, while the B\textsubscript{2g} mode, corresponding to an out-of-plane chalcogen vibration, is comparatively less strain-sensitive but has a pronounced dependence on layer number and interlayer coupling/stacking.\cite{michail2024biaxial} Across all thicknesses, both modes from flakes on graphene consistently show a blue shift of $\approx$1.0–1.5\,\cminv{} compared to the \sio{}-supported flakes. This systematic blue shift indicates the presence of compressive strain in the MOCVD-grown \wse{} on graphene. This finding is consistent with previous studies of \wse{} on graphene, where synchrotron-based grazing-incidence in-plane X-ray diffraction showed a lattice compression of $-$0.19 \% in monolayer \wse{} compared to bulk \wse{}.\cite{nakamura2020spin} Such compression enables a 3 × 3 unit cell of slightly compressed \wse{} to align commensurately with a 4 × 4 graphene lattice, resulting in a lattice mismatch of less than 0.03 \%. \cite{nakamura2020spin} In this case, the compressive strain is preserved even in multilayer \wse{}, indicating that the system does not relax upon increasing thickness.

In contrast, \wse{} flakes on \sio{} likely experience tensile strain, as substrate roughness, limited epitaxial order, and surface charge inhomogeneities allow local relaxation or stretching of the 2D layer.\cite{sahin2013anomalous,lin2021thermal} These effects promote a disordered interface and tensile contributions, leading to Raman redshifts that enhance the frequency difference between \wse{} on graphene and on \sio{}.

Figure~\ref{fig3}c presents the frequency difference between the B\textsubscript{2g} and A\textsubscript{1g}/E\textsubscript{2g} modes, serving as a useful methodology for layer number identification, particularly in cases where absolute peak positions are influenced by substrate effects or local strain variations. On both substrates, this frequency difference decreases with increasing layer number, but \wse{} on graphene consistently shows lower values than on \sio{}, supporting the notion of stronger interface effects in the graphene-substrate system. 
The frequency difference between the A\textsubscript{1g}/E\textsubscript{2g} and B\textsubscript{2g} modes may also be influenced by charge transfer at the \wse{}/graphene interface.\cite{nakamura2020spin} The thickness-dependent blue shift of the graphene G and 2D modes, discussed below, indicates electron transfer from graphene to \wse{}, leading to progressive $n$-type doping of \wse{} and $p$-type doping of graphene with increasing layer number.\cite{zhang2021charge,fates2019probing,das2008monitoring} This charge transfer can further stiffen the \wse{} lattice and contribute to the observed Raman shifts.\cite{fan2020tailoring}

\subsubsection{Interlayer Phonon Modes and Substrate-Dependent Coupling in \wse{}}

\begin{figure}[t!]
\centering
\includegraphics[width=1.0\textwidth]{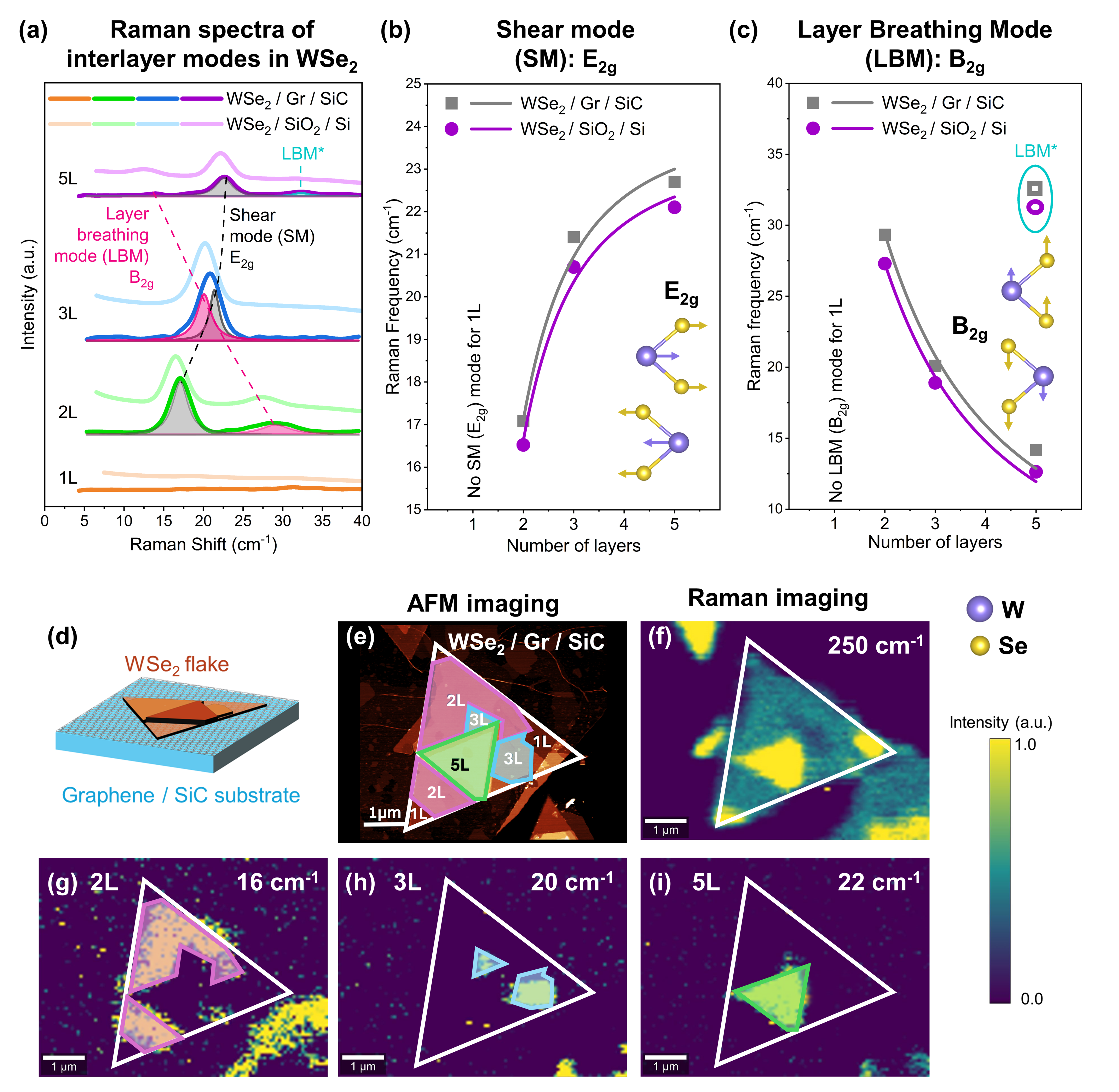}
\caption{\label{fig4} \textbf{Raman and spatial characterization of interlayer vibrational modes in \wse{} on graphene/SiC and \sio{}/Si substrates.}
(a) Low-frequency Raman spectra showing shear (E\textsubscript{2g}) and layer breathing (B\textsubscript{2g}) modes appearing from bilayer \wse{}. Raman spectra were deconvoluted using Lorentzian fitting to extract mode positions. A higher-order breathing mode (LBM*) is observed for 5L \wse{} on graphene. (b) Shear and (c) layer breathing mode frequencies as a function of \wse{} layer thickness on graphene/SiC (gray squares) and on \sio{}/Si (purple circles, data from Blaga et al., \cite{blaga_unveiling_2024}). Solid lines represent fits using the linear chain model (LCM), from which the in-plane ($K_x$) and out-of-plane ($K_z$) interlayer force constants are extracted. (d) Schematic of \wse{}/graphene/SiC sample. (e) AFM image showing layer distribution. (f) Raman map of the 250\,\cminv{} intralayer mode. (g–i) Raman maps of shear modes at 16, 20, and 22\,\cminv{} for 2L, 3L, and 5L regions.}
\end{figure}

Figure~\ref{fig4} presents the analysis of interlayer vibrational modes in \wse{} flakes grown on graphene, focusing on both shear modes (SM) and layer breathing modes (LBM) as a function of layer number. Figure~\ref{fig4}a shows the low-frequency Raman spectra, where both the shear (E\textsubscript{2g}) and breathing (B\textsubscript{2g}) modes become active starting from bilayer \wse{}. To accurately determine the positions of these interlayer modes, the low-frequency spectral region was deconvoluted using Lorentzian peak fitting developed by Dimitrievska et al., and as detailed in Ref. \cite{blaga_unveiling_2024,nielsen2025bazrs3} . The frequency evolution of the shear and breathing modes is shown in Figures~\ref{fig4}b and~\ref{fig4}c. Both modes are systematically blue-shifted for \wse{} on graphene compared with \sio{}. For the shear modes, this is consistent with reduced interlayer spacing or stronger interlayer coupling.\cite{zhang_raman_2013} The breathing-mode shift also indicates stronger out-of-plane coupling. An additional high-frequency breathing mode, LBM*, appears near 32\,\cminv{} for 5L \wse{} on graphene but is absent on \sio{}. We identify this as the third-order breathing mode, LBM\textsubscript{3}, which can become Raman active through complex interlayer force distributions or substrate-induced symmetry breaking.\cite{zhang_raman_2013}

Figures~\ref{fig4}f--\ref{fig4}i map the intensity of the intralayer mode at 250\,\cminv{} and the shear modes at 16, 20, and 22\,\cminv{} for the 2L, 3L, and 5L regions, respectively. Because the shear-mode frequency increases monotonically from 2L to 5L, these maps distinguish the different layer regions and provide a reliable measure of layer number.\cite{zhang_raman_2013}

To quantify the substrate-induced changes in interlayer coupling, we analyze the shear and breathing mode frequencies using the linear chain model (LCM).\cite{tan2012shear,zhang_raman_2013} In this model, each \wse{} monolayer is treated as a rigid mass per unit area $\mu$, coupled to adjacent layers by harmonic springs that represent the effective interlayer restoring forces. Only nearest-neighbor interactions are considered, and each layer is assumed to move as a single unit. We define two independent force constants per unit area: the out-of-plane force constant $K_z$, related to LBM, and the in-plane force constant $K_x$, associated with interlayer SM. Solving the equations of motion for a stack of $N$ coupled layers results in a discrete set of normal modes with characteristic layer-number-dependent frequencies. For the lowest-frequency breathing mode, the LBM frequency is given as:\cite{zhang_raman_2013}

\begin{equation}
\nu_{\mathrm{LBM}}(N) =
\frac{1}{2\pi c}
\sqrt{
\frac{2K_z}{\mu}
\left[
1 - \cos\left(\frac{\pi}{N}\right)
\right]
},
\end{equation}

where $c$ is the speed of light and $\mu$ is the mass per unit area of a \wse{} trilayer. Similarly, the frequency of the SM mode can be calculated as:

\begin{equation}
\nu_{\mathrm{SM}}(N) =
\frac{1}{2\pi c}
\sqrt{
\frac{2K_x}{\mu}
\left[
1 + \cos\left(\frac{\pi}{N}\right)
\right]
}.
\end{equation}

Based on these equations, the LBM frequency decreases with increasing layer number due to reduced restoring force per layer, while the SM frequency increases as the number of coupled layers increases, in agreement with the experimental dependencies shown in Figures~\ref{fig4}b and~\ref{fig4}c.

We fit the experimentally measured LBM and SM frequencies using the above expressions, treating $K_z$ and $K_x$ as fitting parameters and fixing the mass per unit area to $\mu = 6.08 \times 10^{-6}\,\mathrm{kg\,m^{-2}}$, corresponding to the known value for \wse{}.\cite{zhang_raman_2013} The resulting fits match well the experimental data, confirming the validity of the LCM for describing interlayer vibrations in our samples. The obtained interlayer force constants are summarized in Table~\ref{tab:force_constants}. 

\begin{table}[h]
\centering
\caption{Interlayer force constants per unit area calculated using the linear chain model for \wse{} on different substrates.}
\label{tab:force_constants}
\begin{tabular}{lcc}
\hline
\textbf{Substrate} & \textbf{$K_z$ ($10^{19}$ N m$^{-3}$)} & \textbf{$K_x$ ($10^{19}$ N m$^{-3}$)} \\
\hline
\wse{} / Gr / SiC & $9.3 \pm 0.5$ & $3.15 \pm 0.08$ \\
\wse{} / \sio{} / Si & $8.1 \pm 0.3$ & $2.98 \pm 0.05$ \\
\hline
\end{tabular}
\end{table}

Based on the values in Table~\ref{tab:force_constants}, we observe an increase of interlayer coupling for \wse{} grown on graphene compared to \sio{}. The increase in both out-of-plane and in-plane force constants indicates stronger mechanical coupling between adjacent layers of \wse{} grown on graphene compared to \sio{}, with a more pronounced effect observed for the breathing modes, reflecting the higher sensitivity of out-of-plane vibrations to the interfacial environment. This enhancement can be attributed to the atomically flat and chemically inert surface of graphene, which provides a more uniform van der Waals interface and minimizes surface roughness, trapped charges, and interfacial disorder that are typically present in \sio{}. As a result, the \wse{} layers have a more homogeneous interlayer spacing and stronger effective interlayer interactions. These values are also in good agreement with previously reported interlayer force constants for multilayer \wse{}.\cite{zhang_raman_2013} In the following discussion, these interlayer force constants are used to estimate the contribution of modified interlayer coupling to the exciton-energy shifts in multilayer \wse{}.

\subsubsection{Quantification of Charge Transfer at the \wse{}/Graphene Interface from Graphene G and 2D Modes}

\begin{figure}[t!]
\centering
\includegraphics[width=0.7\textwidth]{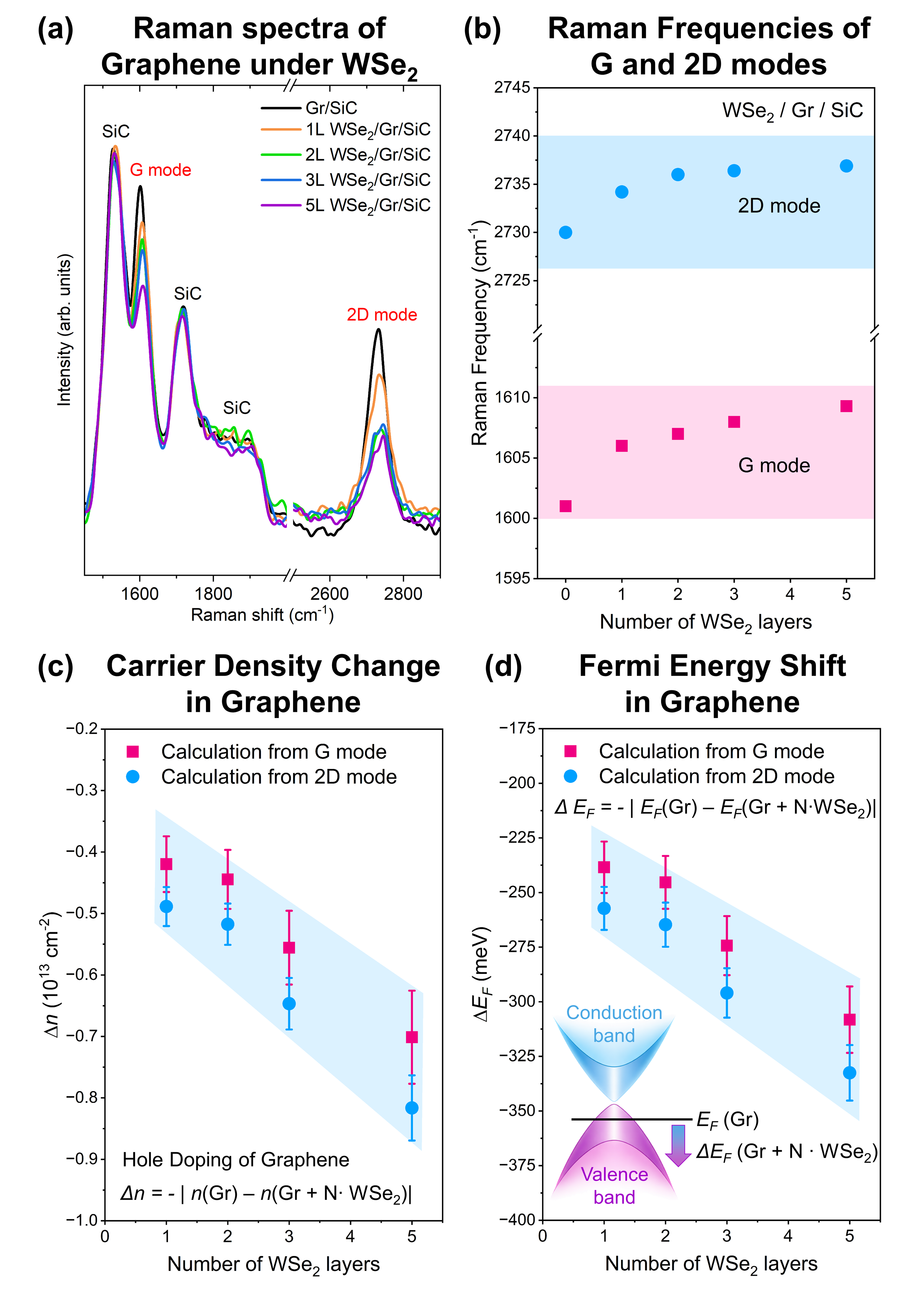}
\caption{
\textbf{Raman analysis of graphene doping induced by WSe$_2$.}
(a) Raman spectra of quasi-freestanding epitaxial graphene on SiC with increasing WSe$_2$ layer coverage (1L to 5L), showing the graphene G and 2D modes together with second-order SiC phonon peaks. 
(b) Evolution of the graphene G and 2D mode frequencies as a function of WSe$_2$ thickness, showing systematic blue shifts consistent with charge transfer. 
(c) Change in graphene carrier density extracted from strain-corrected Raman frequencies of G and 2D modes and using calibration coefficients derived from Das \textit{et al.}, indicating progressive hole doping with increasing WSe$_2$ thickness. 
(d) Corresponding Fermi level shift relative to bare graphene, showing a downward shift of the Fermi level toward the valence band, consistent with increasing p-type doping. 
}
\label{fig5}
\end{figure}

Following the detailed investigation of the \wse{} vibrational modes, we now turn to the analysis of the graphene Raman features under the \wse{} flakes. Figure~\ref{fig5}a shows the Raman spectra of the graphene/SiC substrate as a function of \wse{} layer number, focusing on the G and 2D modes. It is important to note that the positions of these modes in epitaxial graphene on SiC are inherently different from those typically observed in exfoliated graphene on \sio{}. Epitaxial graphene on SiC is known to experience compressive strain because of the significant difference in thermal expansion coefficients between graphene and the SiC substrate,\cite{ferralis2008evidence} resulting in a characteristic blue shift of the G mode (from $\approx$1583\,\cminv{} in unstrained graphene to $\approx$1600–1610\,\cminv{} on SiC) and a blue-shifted, broadened 2D mode centered near $\approx$2730\,\cminv{} \cite{rohrl_raman_2008}. The Raman spectra in Figure~\ref{fig5}a also show two prominent SiC phonon modes at approximately 1520\,\cminv{} and 1715\,\cminv{}, which are consistent with the reported second-order Raman features of 6H-SiC.\cite{zhang_raman_2018}

Both the G and 2D mode frequencies of graphene increase systematically with increasing \wse{} thickness. The G mode shifts from $\approx$1601\,\cminv{} for bare graphene/SiC to $\approx$1609\,\cminv{} beneath the 5-layer \wse{} flake, while the 2D mode increases from $\approx$2730\,\cminv{} to $\approx$2737\,\cminv{}. Such simultaneous blue shifts of the G and 2D modes point to charge transfer and modification of the graphene carrier density.\cite{das2008monitoring,fates2019probing} This interpretation is supported by the correlated evolution of the two modes, shown in Figure~S2 in the Supporting Information, which follows a linear trend with slope $\Delta\omega_{2D}/\Delta\omega_G = 0.85 \pm 0.05$. This value is close to the characteristic slope expected for doping-dominated behavior ($\sim$0.7) and is significantly smaller than the slope associated with purely strain-induced shifts ($\sim$2.2), confirming that charge transfer is the dominant mechanism governing the graphene Raman frequency evolution.\cite{das2008monitoring,sonntag2023charge} At the same time, the deviation of the observed slope from the ideal doping limit indicates that a residual strain contribution is also present. Therefore, to isolate the purely doping-related component of the Raman frequencies, we perform a detailed strain–doping decomposition of the G and 2D mode frequencies, as described in the Supporting Information, enabling quantitative extraction of the carrier density and Fermi level changes associated with charge transfer at the graphene/\wse{} interface.

Using the strain-corrected Raman shifts and the calibration established by Das \textit{et al.}, we calculate the corresponding carrier density changes, shown in Figure~\ref{fig5}c. The results show a monotonic increase in hole concentration with increasing \wse{} thickness, reaching $\Delta n \approx -(0.70 \pm 0.09)\times10^{13}$\,cm$^{-2}$ (G mode) and $\Delta n \approx -(0.82 \pm 0.10)\times10^{13}$\,cm$^{-2}$ (2D mode) for the 5-layer flake. These two independent estimates of hole concentration agree within their uncertainties, with the slightly larger values obtained from the 2D mode reflecting its increased sensitivity to electronic and dielectric environment changes due to its double-resonant origin.\cite{sonntag2023charge} These results confirm that the \wse{} progressively contributes to the increase in the hole density of graphene as the number of \wse{} layers increases.

The corresponding changes in the graphene Fermi energy, calculated from the Raman-derived carrier density using the linear dispersion relation of graphene (see Supporting Information) and summarized in Figure~\ref{fig5}d, show a systematic additional downward shift of the graphene Fermi level upon \wse{} deposition.  Quasi-free-standing graphene on hexagonal SiC is already expected to be intrinsically \emph{p-type} even before \wse{} growth, due to spontaneous polarization of the SiC substrate.\cite{ristein2012origin} Ristein \textit{et al.} report that hydrogen-intercalated quasi-free-standing graphene on 6H-SiC(0001) shows a robust p-type conductivity with the Fermi level $\sim$0.30\,eV below the Dirac energy and hole densities of order $5.5\times10^{12}$\,cm$^{-2}$.\cite{ristein2012origin} In this regard, the values extracted here indicate that adding \wse{} layers drives graphene \emph{further} into the p-doped regime, with the relative shift (referenced to bare graphene) increasing in magnitude from $\approx -240$\,meV for 1L \wse{} to $\approx -330$\,meV for the 5L flake.

This progressive increase in p-type doping of graphene is consistent with interfacial charge transfer set by the band alignment between graphene and \wse{}. As shown by Pan \textit{et al.},\cite{pan2022exciton} the work function of graphene on SiC ($\approx$4.3\,eV)\cite{mammadov2017work} exceeds the electron affinity of monolayer \wse{} (3.7--3.9\,eV), making electron transfer from graphene into \wse{} energetically favorable. Additional \wse{} deposition further depletes electrons from an already p-doped graphene layer, increasing the hole density and shifting $E_F$ downward relative to the bare-graphene reference. This interpretation is also supported by time-resolved measurements by Rieland \textit{et al.},\cite{rieland_ultrafast_2024} who demonstrated ultrafast electron transfer from graphene to \wse{} and transient population of the \wse{} conduction band, directly corroborating interfacial charge transfer in graphene/\wse{} heterostructures.

Additionally, Figure~S4 in the Supporting Information shows a progressive decrease in G mode intensity with increasing \wse{} thickness. This behavior can be attributed to two complementary mechanisms. First, increased carrier density in graphene modifies electron–phonon coupling and leads to Pauli blocking, which reduces the Raman scattering efficiency of the G mode.\cite{rao2019Pauli,chen2011Pauli,casiraghi2009Pauli} Second, the \wse{} overlayer introduces additional optical attenuation, partially absorbing both the incident laser light and the scattered Raman signal.\cite{gu2019layer}

\subsection{Photoluminescence characterization of MOCVD-Grown \wse{} on Graphene/SiC substrate}\label{sec4}

\begin{figure}[t!]
\centering
\includegraphics[width=0.72\textwidth]{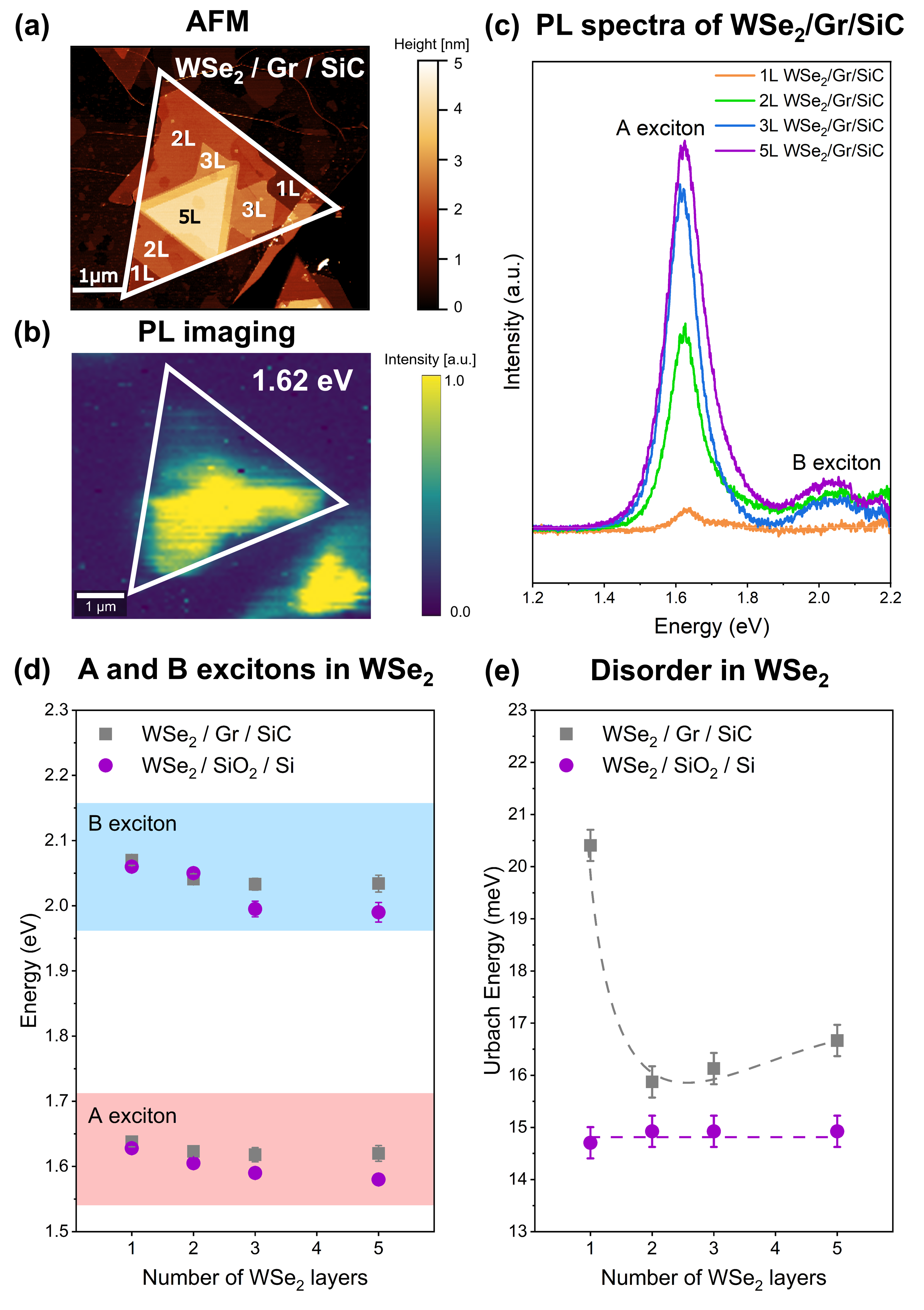}
\caption{\label{fig6} \textbf{Photoluminescence of \wse{} on Graphene/SiC substrate. }
(a) AFM image of a MOCVD-grown \wse{} flake on bilayer graphene/SiC. (b) Spatially resolved photoluminescence (PL) imaging at 1.62\,eV highlights the strong layer-dependent PL intensity, with significant quenching observed in the monolayer region. (c) Average PL spectra for 1L, 2L, 3L, and 5L \wse{} flakes, showing a clear increase in PL intensity with layer number and the evolution of both A and B excitonic features. (d) Extracted A and B exciton energies as a function of \wse{} layer thickness on graphene/SiC (gray squares) and on \sio{}/Si (purple circles, data from Ref.~\cite{blaga_unveiling_2024}) (e)Urbach energies obtained by fitting the exponential absorption tail of the absorbance spectra derived from the PL data for \wse{} on graphene/SiC and \sio{}/Si. The dashed lines are guides to the eye.}
\end{figure}

PL measurements of \wse{} show a strong dependence of the optical emission on both the number of layers and the underlying substrate. Figure~\ref{fig6} summarizes the spatially resolved and layer-dependent PL properties of MOCVD-grown \wse{} on bilayer graphene/SiC. The AFM image in Figure~\ref{fig6}a shows the 1L, 2L, 3L, and 5L regions of the investigated flake, while the PL intensity map at 1.62,eV in Figure~\ref{fig6}b shows the corresponding spatial distribution of the emission. The average PL spectra extracted from these regions are presented in Figure~\ref{fig6}c, followed by the layer-dependent A- and B-exciton energies in Figure~\ref{fig6}d and the effective Urbach energies in Figure~\ref{fig6}e. For comparison, the exciton and Urbach energies of exfoliated \wse{} on \sio{}/Si are also included. The exciton energies for the \sio{}/Si-supported flakes were reported previously in Ref.~\cite{blaga_unveiling_2024}.

The PL spectra measured using 488\,nm excitation at room temperature show a pronounced increase in emission intensity with increasing \wse{} layer number. The monolayer shows only weak emission around the A-exciton energy, in agreement with the strongly suppressed intensity observed in the spatial PL map. A substantial increase is already observed for 2L \wse{}, followed by a further increase for the 3L and 5L regions. In addition to the dominant A-exciton emission near 1.62\,eV, all PL spectra contain a weaker and broader B-exciton feature between approximately 2.0 and 2.1\,eV. The low PL intensity of the monolayer is consistent with the efficient interfacial charge- and energy-transfer processes previously reported for directly contacted \wse{}/graphene heterostructures.\cite{lorchat2020filtering,rieland_ultrafast_2024,froehlicher2018charge} The progressive recovery of the PL intensity with increasing layer number is also consistent with the rapid decrease of nonradiative energy transfer to graphene as the emitting region becomes spatially separated from the interface.\cite{swathi2009distance,federspiel2015distance} Although multilayer \wse{} has an indirect band gap, measurable emission can still arise from phonon-assisted recombination.\cite{brem2020phonon}

The A- and B-exciton energies were obtained by fitting each PL spectra with a single Gaussian function per exiton, as shown in Figure S5 of the Supporting Information. No additional trion contributions were included in the fitting of the A-exciton region because the weak PL intensity, especially monolayer \wse{} on graphene, did not allow the neutral-exciton and trion contributions to be resolved reliably.

The A- and B-exciton energies show very different thickness dependencies on graphene/SiC and \sio{}/Si. For \wse{} on graphene/SiC, both excitonic transitions show relatively small layer-dependent shifts, varying over total energy ranges of approximately 30\,meV for the A exciton and 40\,meV for the B exciton between 1L and 5L. In comparison, the corresponding shifts on \sio{}/Si are larger, reaching around 50\,meV for the A exciton and 70\,meV for the B exciton. The transition energies of the monolayers on the two substrates are comparable, differing by only approximately 10\,meV. From 2L onward, however, both the A- and B-exciton transitions occur systematically at higher energies for \wse{} on graphene/SiC, and the difference between the two substrates increases with layer number. This shows that graphene tends to pin both the A- and B-exciton energies. The origin of this pinning is discussed in the Discussion section through a quantitative analysis of the exciton shifts in relation to the observed compressive strain, changes in interlayer distance, electronic screening and charge transfer.

\subsubsection{Urbach energy and Disorder in \wse{}}

To investigate how the graphene interface and increasing layer number affect the optical band-edge broadening in \wse{}, we determine the effective Urbach energy. The Urbach energy characterizes the exponential broadening of the band-edge absorption tail and reflects the distribution of local band-edge energies arising from structural defects, strain variations, and spatial inhomogeneities in the electronic environment.\cite{ugur2022life,linke2026cubic,stutz2022stoichiometry} The effective Urbach energies were determined from the low-energy tails of the PL spectra following the generalized Planck relation.\cite{ugur2022life} In the low-energy spectral range, the PL intensity can be approximated as

\begin{equation}
\mathrm{PL}(E)\propto A_{\mathrm{PL}}(E)E^{2}
\exp\left(-\frac{E}{k_{\mathrm{B}}T}\right),
\end{equation}

where $A_{\mathrm{PL}}(E)$ is the absorptance derived from the PL spectrum, $E$ is the photon energy, $k_{\mathrm{B}}$ is the Boltzmann constant, and $T$ is the measurement temperature. The corresponding PL-derived absorptance is then calculated as

\begin{equation}
A_{\mathrm{PL}}(E)\propto
\frac{\mathrm{PL}(E)}{E^{2}}
\exp\left(\frac{E}{k_{\mathrm{B}}T}\right).
\end{equation}

Within the Urbach-tail region, the logarithm of the absorptance varies linearly with photon energy, and the effective Urbach energy, $E_{\mathrm{U}}$, is obtained from

\begin{equation}
\frac{d\left[\ln A_{\mathrm{PL}}(E)\right]}{dE}
\simeq \frac{1}{E_{\mathrm{U}}(T)}.
\end{equation}

The corresponding linear fits are shown in Figure~S6 of the Supporting Information.

As shown in Figure~\ref{fig6}e, the effective Urbach energy is highest for monolayer \wse{} on graphene/SiC, at around 20\,meV, decreases to around 16\,meV for 2L, and then increases slightly toward 5L. By comparison, \wse{} on \sio{}/Si shows nearly layer-independent values of approximately 15\,meV. The values measured here are comparable to the sharp optical edge reported for a monolayer \wse{}/MoS$_2$ heterobilayer, for which an inverse Urbach-tail slope of approximately 30\,meV per decade was extracted from PL, corresponding to $E_{\mathrm{U}}\approx13$\,meV when expressed using a natural-logarithmic slope.\cite{fang2014strong} 

Since the two sample systems differ in both the substrate and preparation method, the higher value for monolayer \wse{} on graphene/SiC cannot be assigned to growth-related defects alone. Instead, it is consistent with stronger broadening in the layer directly coupled to graphene, arising from the combined effects of charge transfer, compressive strain, and variations in the local environment. The sharp decrease from 1L to 2L suggests that this interfacial contribution is rapidly screened by the addition of a second \wse{} layer. The slight increase in Urbach Energy from 2L to 5L can be related to increasing structural disorder within the MOCVD-grown multilayer stack. Each additional layer introduces another interlayer interface, where imperfect alignment can result in stacking faults, small rotational misalignments, local variations in interlayer spacing, and strain gradients.\cite{shearer2017complex} Screw-dislocation-driven vertical growth may also produce increasingly complex stacking sequences in multilayer CVD-grown \wse{}.\cite{chen2014screw} These structural variations broaden the distribution of local band-edge and excitonic transition energies and therefore increase the effective Urbach energy. In our samples, this interpretation is consistent with the changes in the interlayer Raman modes and the additional high-frequency breathing mode observed for 5L. However, the simultaneous evolution of charge transfer, screening, strain, and interlayer coupling may also contribute to this trend. On the other hand, the nearly constant $E_{\mathrm{U}}$ values of exfoliated \wse{} on \sio{}/Si suggest that increasing thickness does not introduce a comparable degree of additional disorder. This is consistent with the exfoliated flakes retaining the ordered stacking of the parent crystal, without the successive growth interfaces and stacking variations that can develop during MOCVD growth.

\section{Discussion}

The reduced thickness dependence of the A- and B-exciton energies can be understood quantitatively from the combined effects of strain, dielectric screening, charge transfer, and interlayer coupling. Between 1L and 5L, the A- and B-exciton energies redshift by approximately 30 and 40\,meV, respectively, on graphene/SiC, compared with approximately 50 and 70\,meV on \sio{}/Si. Graphene therefore suppresses the layer-dependent redshift by around 20\,meV for the A exciton and 30\,meV for the B exciton. Compressive strain can account for a considerable part of this difference. Epitaxial \wse{} on graphene has previously been reported to exhibit a lattice compression of approximately $-0.19$\%.\cite{nakamura2020spin} Experimental strain coefficients for the A exciton range from approximately 48\,meV/\% under uniaxial strain to $118\pm6$\,meV/\% under biaxial strain.\cite{aslan2018strain,kumar2024strain} Consequently, a compression of this magnitude is expected to induce an A-exciton blueshift of approximately 9--22\,meV. Since the Raman shifts indicate that the compressive strain is largely maintained from 1L to 5L, strain mainly produces an upward shift of the exciton energies and can only partly explain their reduced thickness dependence.

Electronic screening can make a contribution of several tens of meV to the optical transition energy. For monolayer \wse{}, a local high-$\kappa$ HfO$_2$ overlayer was reported to produce an A-exciton redshift of approximately 17\,meV, while comparisons across different low- and high-$\kappa$ dielectric environments showed optical-gap shifts of up to 37\,meV.\cite{gauriot2024direct,borghardt2017engineering} A screening contribution of the order of 20--40\,meV is therefore reasonable for the A-exciton energy. Although graphene can renormalize the quasiparticle gap and exciton binding energy by more than 100\,meV, these two changes largely compensate in the optical transition.\cite{raja_coulomb_2017} As the emitting states become progressively separated from graphene with increasing \wse{} thickness, the reduced influence of graphene screening can oppose part of the intrinsic layer-dependent redshift and contribute to the observed exciton pinning. 

At the same time, the increasing charge transfer from graphene to \wse{} introduces an additional carrier-induced renormalization of the excitonic transitions. The graphene Raman modes indicate a transferred charge density of approximately $(0.70$--$0.82)\times10^{13}$\,cm$^{-2}$ for 5L. However, this value represents the carrier-density change in graphene and cannot be directly equated with the electron density in \wse{}, whose distribution across the individual layers is not known and is expected to be strongest near the interface. Room-temperature electrochemical gating measurements of monolayer \wse{} showed that the neutral A exciton blueshifts by only $2\pm1$\,meV at an electron density of approximately $1.0\times10^{13}$\,cm$^{-2}$, owing to the partial compensation between band-gap renormalization and the reduction in exciton binding energy.\cite{morozov2021room} Charge transfer is therefore expected to shift the neutral exciton in our samples by only a few meV, considerably less than the contributions from strain and dielectric screening. Because charge transfer is expected to be strongest in the \wse{} layer directly adjacent to graphene, this interfacial layer should also experience the fastest nonradiative relaxation into graphene and therefore contribute only weakly to the measured PL. The observed multilayer emission is therefore expected to originate mostly from the upper \wse{} layers, which are farther from graphene and which are probably less strongly doped.

Finally, the stronger interlayer coupling measured for \wse{} on graphene/SiC provides a smaller but non-negligible contribution to the excitonic response. The out-of-plane and in-plane interlayer force constants increase by approximately 15\% and 6\%, respectively, relative to \wse{} on \sio{}/Si. Within a simplified 12--6 Lennard--Jones description, for which $K_z\propto d^{-8}$, the measured increase in $K_z$ corresponds to an estimated reduction of the interlayer distance from approximately 6.48 to 6.37\,\AA, or $\Delta d=0.11\pm0.05$\,\AA\, as shown in the Supporting Information.\cite{devries2023thermal,schutte1987crystal} This represents a relative compression of approximately 1.7\%. Using the reported bulk modulus of \wse{}, $B_0\approx72$\,GPa, this compression corresponds to an effective pressure scale of approximately 1.2\,GPa.\cite{selvi2008high} Pressure-dependent measurements of trilayer \wse{} reported an increase in the layer-hybridization strength of $3.5\pm0.2$\,meV/GPa, implying a change of approximately 4\,meV for the estimated compression.\cite{zhao2026pressure} This value represents a change in the interlayer-hybridization energy rather than a direct shift of the A- or B-exciton transition. Nevertheless, it indicates that the contribution of the modified interlayer coupling is likely limited to a few meV and is therefore smaller than the tens-of-meV contributions associated with strain and dielectric screening.

Taken together, these estimates indicate that the exciton pinning is governed primarily by the competition between the intrinsic thickness-dependent redshift of \wse{} and the progressively weaker screening influence of graphene as the optically active states move farther from the interface. The reduction in graphene screening with increasing layer number can produce a relative blueshift of several tens of meV, thereby compensating a substantial part of the redshift normally observed from 1L to 5L. Compressive strain contributes an additional A-exciton blueshift of approximately 9--22\,meV and helps maintain the transition energies at systematically higher values than on \sio{}/Si. However, because the Raman results indicate that the strain remains approximately constant across the investigated thickness range, strain primarily determines the overall energy offset rather than the layer-dependent pinning itself. Charge transfer and interlayer coupling are expected to shift the exciton transition by only a few meV. These effects therefore provide secondary corrections rather than the dominant compensation mechanism. The suppression of the layer-dependent redshift by approximately 20\,meV for the A exciton and 30\,meV for the B exciton is thus most consistently explained by distance-dependent electronic screening, reinforced by a persistent compressive-strain-induced blueshift and smaller contributions from charge transfer and enhanced interlayer coupling. Although the numerical values remain approximate and are not strictly additive, they establish a clear hierarchy among the different mechanisms responsible for the observed pinning. Future time-resolved, gate-dependent, and direct structural measurements will be important for separating these contributions and establishing design rules for optoelectronic devices.

\section{Conclusions}

In conclusion, the direct comparison of MOCVD-grown \wse{} on graphene/SiC with exfoliated \wse{} on \sio{}/Si establishes graphene as an active interfacial component rather than a passive substrate. Raman spectroscopy shows persistent compressive strain, progressive electron transfer from graphene to \wse{}, and enhanced interlayer coupling, while PL measurements show strong monolayer quenching followed by emission recovery in thicker regions. Despite these pronounced interfacial effects, the A- and B-exciton energies shift by only approximately 30 and 40\,meV between 1L and 5L on graphene, compared with approximately 50 and 70\,meV on \sio{}/Si. The quantitative analysis indicates that this pinning originates primarily from the decreasing dielectric influence of graphene with increasing \wse{} thickness, while compressive strain maintains a higher overall transition energy, and charge transfer and interlayer hybridization provide smaller corrections. The effective Urbach energies further distinguish the strong interfacial broadening of the monolayer from the gradual development of structural disorder in the MOCVD-grown multilayers. These results show that controlling graphene carrier density, interlayer spacing, strain, and stacking during growth could provide a route to independently tune exciton energies and emission efficiency in scalable van der Waals heterostructures. Future work should explore how this thickness-insensitive excitonic response can be translated into electrically tunable \wse{}/graphene photodetectors, light-emitting devices, and modulators with improved device-to-device reproducibility.

\clearpage


\section{Methods}

\subsection*{Sample preparation}

\subsubsection*{MOCVD \wse{} samples on Gr/SiC substrates}
Our \wse{} samples are grown via metal organic chemical vapor deposition on few-layer quasi-freestanding epitaxial graphene supported by a conductive silicon carbide (c-SiC) substrate. Intercalation of the graphene layers with hydrogen prior to \wse{} growth terminates the dangling silicon bonds, providing homogeneous electrostatic background. Further details of the sample growths procedure were previously reported along references. \cite{bobzien2025layer, briggs2016realizing}

\subsubsection*{Exfoliated \wse{} samples on SiO\textsubscript{2}/Si substrates}

Reference multilayer \wse{} flakes were obtained by mechanically exfoliating bulk \wse{} crystals (HQ Graphene) using adhesive tape and subsequently transferred onto silicon substrates with a 285\,nm SiO\textsubscript{2} layer (NOVA Electronic Materials) via the same method. Detailed characterization and properties of these flakes are presented in Ref. \cite{blaga_unveiling_2024}. The optical microscopy and AFM image of the flake used is given in Figure S1 in the Supporting Information. For the purpose of this study, and to obtain reference Raman spectra for each layer, we remeasured all layers of the flake using a 488 nm laser under the same conditions as those applied to the MOCVD-grown \wse{}.

\subsection*{Experimental methods}

\subsubsection*{Raman and photoluminescence (PL) Spectroscopy}

Raman and photoluminescence (PL) measurements were performed using a WITec Alpha 300 R confocal Raman microscope in backscattering geometry. Raman and PL spectra were acquired using a 488\,nm excitation laser. The laser beams were focused onto the sample through a high-magnification microscope objective, resulting in a spot size of approximately 700\,nm for the 488\,nm laser. The radiant output power at the sample surface was maintained around 150\,\textmu W. To ensure that the chosen excitation power did not induce heating or damage, a power-dependent study was conducted by measuring spectra at a fixed point on the sample under increasing laser power. Any shifts in peak position, broadening, or emergence of new spectral features were monitored. The maximum power at which no such changes were observed (150\,\textmu W) was used for all measurements. The Raman signal was analyzed using a 300\,mm focal-length spectrometer equipped with an 1800\,g/mm grating and a thermoelectrically cooled CCD, while the PL spectra were acquired using a 150\,g/mm grating. All Raman peak positions were calibrated using the characteristic silicon peak at 520\,cm\textsuperscript{-1}.

Each Raman and PL hyperspectral maps were acquired over an area of $7\times7$\,\textmu m$^{2}$ using a $100\times100$ point grid, corresponding to 10\,000 spectra per measurement. The layer-specific spectra presented in the figures were obtained by grouping the mapped data into clusters corresponding to the different \wse{} thicknesses and averaging all spectra within each cluster, resulting in an improved signal-to-noise ratio and reduced influence of local spectral variations.

\subsubsection*{AFM characterization}
Atomic Force Microscopy (AFM) measurements were carried out using a Bruker Dimension Icon 3 microscope equipped with NanoScope software. The samples were imaged in tapping mode utilizing Bruker ScanAsyst-Air probes. This imaging mode was selected to minimize tip-sample interaction forces and to prevent potential damage to the flakes during scanning. The acquired AFM data were further analyzed and processed using Gwyddion software.


\clearpage


\section*{Supplementary information}
Supporting Information is available at XXX.

\section*{Acknowledgements}
We acknowledge fruitful discussions with Prof. Michel Calame and Prof. Greta Patzke.

\section*{Declarations}
$\dag$ These authors contribute equally to this work.

\subsection{Funding} 
LB and BS appreciate funding by the European Research Council (ERC) under the European Union's Horizon 2020 research and innovation program (Grant agreement No. 948243). Funding for DCF and JAR is through the National Science Foundation EEC-2113864 and ECCS-2202280. C.D. and J.A.R. acknowledges the National Science Foundation, award numbers (DMR-2039351). MD acknowledges Zonta Foundation for Career Fund. For the purpose of Open Access, the author has applied a CC BY public copyright license to any Author Accepted Manuscript version arising from this submission.

\subsection{Conflict of interest/Competing interests} 
The authors declare no conflict of interest. 

\subsection{Data availability}
The data supporting this study is available at 10.5281/zenodo.17079939.

\subsection{Author contribution}
MD and LB conceived the experiment. DCF, LSL, and CD grew the samples. LH and ALA did the Raman measurements, ALA did the AFM measurements, and LH and MD performed the PL measurements. LH curated the data and carried out the initial data analysis. MD performed the quantitative analysis, integrated the different datasets, and developed the overall interpretation of the results. MD, LB, JR, MC, and BS supervised the students involved in the study. MD and LD wrote the manuscript with feedback from all authors.

\clearpage

\bibliographystyle{rsc}
\bibliography{ref_2}

\end{document}